\documentstyle[psfig]{cupconf}

\ifoldfss
\else
  \ifnfssone
    \newmathalphabet{\mathit}
      \addtoversion{normal}{\mathit}{cmr}{m}{it}
      \addtoversion{bold}{\mathit}{cmr}{bx}{it}
    \newmathalphabet{\mathcal}
      \addtoversion{normal}{\mathcal}{cmsy}{m}{n}
    \else
    \ifnfsstwo
    \fi
  \fi
\fi

\def\hexnumber#1{\ifcase#1 0\or1\or2\or3\or4\or5\or6\or7\or8\or9\or
 A\or B\or C\or D\or E\or F\fi }

\makeatletter
\ifx\CUP@mtlplain@loaded\undefined
\else
\fi
\makeatother
 \makeatletter
 \ifx\CUP@mtlplain@loaded\undefined
   \font\tenbmi=cmmib10 at 10pt
   \font\sevenbmi=cmmib10 at 7pt
   \font\fivebmi=cmmib10 at 5pt

   \newfam\bmifam
   \textfont\bmifam=\tenbmi
   \scriptfont\bmifam=\sevenbmi
   \scriptscriptfont\bmifam=\fivebmi
   
 \fi
 \makeatother
\ifnfsstwo

\fi
\ifnfssone

\fi
\ifoldfss

\fi

\mathchardef\varLambda="0103

\makeatletter
\ifx\CUP@mtlplain@loaded\undefined
\else
\fi
\makeatother
\makeatletter
\ifx\CUP@mtlplain@loaded\undefined
  \font\tenbms=cmbsy10
  \font\sevenbms=cmbsy10 at 7pt
  \font\fivebms=cmbsy10 at 5pt
  \newfam\bmsfam
  \textfont\bmsfam=\tenbms
  \scriptfont\bmsfam=\sevenbms
  \scriptscriptfont\bmsfam=\fivebms

  \edef\bsy@{\hexnumber\bmsfam}
  \mathchardef\bnabla="0\bsy@72
\fi
\makeatother
\title[Dark Matter and Dark Energy] {The Standard Model, Dark Matter, and Dark Energy:  From the Sublime to the Ridiculous}

\author[Lawrence M. Krauss ]%
{LAWRENCE M. KRAUSS
}

\affiliation{Departments of Physics and Astronomy, Case Western Reserve University, Cleveland OH}

\begin{document}
\ifnfssone
\else
  \ifnfsstwo
  \else
    \ifoldfss
      \let\mathcal\cal
      \let\mathrm\rm
      \let\mathsf\sf
    \fi
  \fi
\fi

\maketitle

\begin{abstract}
The Standard Model of cosmology of the 1980's was based on a remarkable interplay of ideas from particle theory, experiment and astrophysical observations.  That model is now dead, and has been replaced by something far more bizarre.   Interestingly, the aspect that has survived involves perhaps the most exotic component: dark matter that dominates the gravitational dynamics of all galaxies, and appears to be composed of a sea of new weakly interacting elementary particles.   But this sea of dark matter appears to play second fiddle to an unknown energy density that appears to permeate all of space, causing the expansion of the Universe to accelerate.  We are left with many more questions than answers, and our vision of the future of the Universe has completely changed. 

\vspace{0.5pc}
\noindent{ (Lectures Given at the XIV Canary Islands Winter School in Astrophysics 2002: Dark Matter and Dark Energy in the Universe. Nov 2002 To Appear in the Proceedings)}
\end{abstract}

\firstsection 
\section{Introduction: Why Cosmology?}

Astrophysics and Cosmology involve observations, not experiments, and thus are instrinsically suspect.  Nevertheless, we have learned over the past 30 years that the universe provides a laboratory for exploring fundamental physics that in many cases exceeds the reach of any terrestrial laboratory.

Consider the following:

(i) Energy:  The center of mass energy of accelerators on Earth has increased by a factor of 1000, from 1 GeV to 1 TeV in the thirty years since 1975.  As a result, we have been able to probe the nature of quantum chromodynamics, and the electroweak theories with unprecedented sensitivity, and we hope we are on the threshold of discovering experimental evidence that will shed light on the origin of mass.  Many theorists expect that this may involve the discovery of supersymmetry.    At the same time, we have learned that studying the light element abundances in the Universe sheds insight into Big Bang Nucleosynthesis in a way that directly constrains particle physics up to regime of the QCD phase transition, while the physics of supernovae is dominated by the weak interactions of neutrinos, and may even be sensitive to particles that cannot be probed in accelerators, such as axions.   The highest energy cosmic rays impacting upon the Earth interact with center of mass energies that exceed those that will be accessible at the LHC.  The physics of dark matter may reveal evidence of supersymmetry before it is probed in accelerators, while understanding the origin of baryons in the universe has constrained both the electroweak theory, and possible grand unified theories.  

(ii) Cross Section and Target Mass:  Accelerators probe cross sections in the range down to about
$10^{-38} cm^2$, with targets that are generally less than a kilometer across.  Astrophysics on the other hand allows us to probe processes that involve cross sections well below $10^{-42} cm^2$, and with targets in excess of 10,000 kilometers across.

As a result, astrophysics and cosmology provide great discovery potential.  Because of the limitations of observational uncertainties, however, terrestrial experiments are required if we are to move beyond discovery to exploring the fundamental details of nature at the smallest scales.    In these lectures, I will describe the current observational situation in cosmology, and the theoretical questions that may shed light on unraveling the nature of dark matter, and dark energy.   

\section{The Standard Model and Cosmological Observables}

As early as a decade ago, the uncertainties in the measurement of cosmological parameters was such that few definitive statements could be made regarding cosmological models.  That situation has changed completely.  Instead all cosmological observables have now converged on a single cosmological model.  In this section I will review our present picture.

First, some basic background.  The Universe as we observe it is isotropic, homogeneous, and expanding.  I shall assume that readers are familiar with these basic features, which are 
determined observationally in reverse order, by measuring redshift-distance relations, by examining the number counts of galaxies, and by observations of the cosmic microwave background radiation.  
I shall discuss the measurement of the expansion rate, and of the cosmic microwave background in some detail shortly.   I will briefly 
mention here that one can show, and you can find in any basic textbook on cosmology, that in a flat Euclidean Universe measuring the number of galaxies as a function of magnitude $m$ that if $N(<m) \approx 10^{0.6m}$ then the underlying distribution is basically homogeneous.   Remarkably this relation holds out to significant distances even in our expanding universe today.  

I am also not going to review here the basic features of FRW cosmology, and will assume the reader is familiar with Einstein's Equations for the evolution of the scale factor an expanding isotropic homogeneous universe, which relate the expansion rate, given by the Hubble Constant, $H$ to the density and curvature, where the ratio of the actual matter density today to that required for a flat universe (given by $\rho_c = 3 H^2/8 \pi G$)  is given by the parameter $\Omega_m$.

Within the framework of an isotropic homogeneous expanding universe, there are a finite set of fundamental observables. 
It seems reasonable to divide this into three subsections, Space, Time, and Matter.   Specifically, I shall concentrate on the following:
\\

{\bf Space:}
\begin{itemize}
\item Expansion Rate
\item Geometry
\end{itemize}

{\bf Time:}
\begin{itemize}
\item Age of the Universe
\end{itemize}

{\bf Matter:}
\begin{itemize}
\item Baryon Density
\item Large Scale Structure
\item Matter Density
\item Equation of State
\end{itemize}

\subsection{Space: The Final Frontier:}
\subsubsection{The Hubble Constant}
Arguably the most important single parameter describing the physical universe today is the Hubble Constant.    Since the discovery in 1929 that the Universe is expanding, the determination of the rate of expansion dominated observational cosmology for much of the rest of the 20th century.
The expansion rate,
given by the Hubble Constant, sets the overall scale for most other
observables in cosmology. 

The big news, if any, is that by the end of the 20th century, almost all measurements have converged on a single range for this all important quantity. (I say {\it almost} all, because to my knowledge Alan Sandage still believes the claimed limits are incorrect ( \cite{sandage}). )

Recently, the Hubble Space Telescope Key Project has announced its final results. 
This is the largest scale endeavor carried out over the past decade 
with a goal of achieving a 10 $\%$ absolute uncertainty in the Hubble
constant.   The goal of the project has been to use Cepheid luminosity
distances to 25 different galaxies located within 25 Megaparsecs in order
to calibrate a variety of secondary distance indicators, which in turn
can be used to determine the distance to far further objects of known
redshift.  This in principle allows a measurement of the
distance-redshift relation and thus the Hubble constant on scales where
local peculiar velocities are insignificant.  The five distance
indicators so constrained are:  (1) the Tully Fisher relation,
appropriate for spirals, (2) the Fundamental plane, appropriate for
ellipticals, (3) surface brightness fluctuations, and (4) Supernova Type
1a distance measures, and (5) Supernovae Type II distance measures.  

The Cepheid distances obtained from the HST project include a larger LMC sample to 
calibrate the period-luminosity relation, a new photometric calibration, and correctdions for 
metallicity.  As a result they determined a new LMC distance modulus, of $\mu_o = 18.50 \pm 0.10$ mag.  The number of Cepheid calibrators used for the secondary measures include 21 for the
Tully-Fisher relation, and 6 for each of the Type Ia and surface fluctuation measures.

The HST-Key project reported  measurements 
for each of these methods is presented below (\cite{hst2}).  (While I
shall adopt these as quoted, it is worth pointing out that some critics
have stressed that this involves utilizing data obtained
by other groups, who themselves sometimes report different values of $H_0$).  The first quoted uncertainty is statistical, the second is systematic (coming from such things as LMC zero point measurements, photometry, metallicity uncertainties, and remnant bulk flows).
$$ H_O^{TF} = 71 \pm 3 \pm 7  $$
$$ H_O^{FP} = 82 \pm 6 \pm 9  $$
$$ H_O^{SBF} = 70 \pm 5 \pm 6  $$
$$ H_O^{SN1a} = 71 \pm 2 \pm 6  $$
$$H_O^{SNII}= 72 \pm 9 \pm 7  $$

On the basis of these results, the Key Project reports a weighted average value:

$$ H_O^{WA} =72 \pm 3  \pm 7 \ km s^{-1} Mpc ^{-1}  ( 1 \sigma)  $$

\noindent{and a final combined average of}

$$ H_O^{WA} =72 \pm 8 \ km s^{-1} Mpc ^{-1}  ( 1 \sigma)  $$.

The Hubble Diagram obtained from the HST project (\cite{hst2}) is reproduced as figure 1.

\begin{figure}
\centerline{\psfig{figure=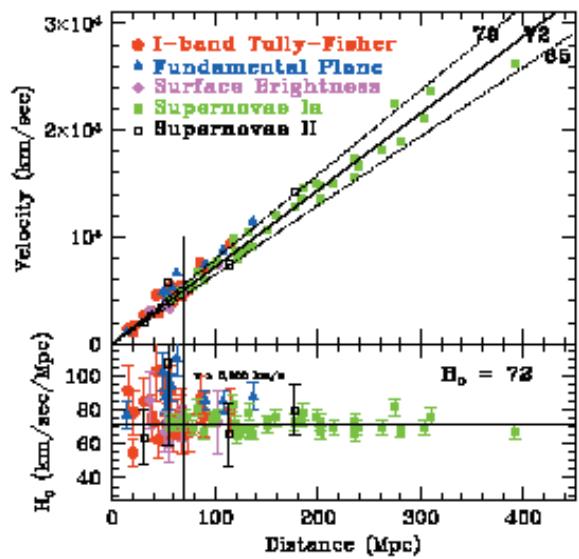,width=9.0cm}}
\label{hubble}
\caption[]{HST Key Project Hubble Diagram}
\end{figure}
 
In the weighted average quoted above, the dominant contribution to 
the $11\%$ one sigma error comes from an overall uncertainty in the
distance to the Large Magellanic Cloud.   If the Cepheid Metallicity 
were shifted within its allowed  4$\%$ uncertainty range, the best fit
mean value for the Hubble Constant from the HST-Key project would shift
downard to  $68 \pm 6$.

\vskip 0.2in
\noindent{\bf \it S-Z Effect:}

The Sunyaev-Zeldovich effect results from a shift in the spectrum of 
the Cosmic Microwave Background radiation due to scattering of the
radiation by electgrons as the radiation  passes through intervening
galaxy clusters on the way to our receivers on Earth.  Because the
electron temperature in Clusters exceeds that in the CMB, the radiation
is systematically shifted to higher frequencies, producing a deficit in
the intensity below some characteristic frequency, and an excess above
it.  The amplitude of the effect depends upon the Thompson scattering
scross section, and the electron density, integrated over the photon's
path:

$$
{\rm SZ} \approx \int{ \sigma_T n_e dl}
$$

At the same time the electrons in the hot gas that dominates the 
baryonic matter in galaxy clusters also emits X-Rays, and the overall
X-Ray intensity is proportional to the {\it square} of the electron
density integrated along the line of sight through the cluster:

$$
{ \rm X-Ray}  \approx \int{n_e^2 dl}
$$

Using models of the cluster density profile one can then use the the
differing dependence on $n_e$ in the two integrals above to extract 
the physical path-length through the cluster.  Assuming the radial
extension of the cluster is approximately equal to the extension across
the line of sight one can compare the physical size of the cluster to the
angular size to determine its distance.  Clearly, since this assumption
is only good in a statistical sense, the use of S-Z and X-Ray
observations to determine the Hubble constant cannot be done reliably on
the basis of a single cluster observation, but rather on an ensemble.  

A recent preliminary analysis of several clusters (\cite{sz})
yields:  

$$
H_0^{SZ} = 60 \pm 10 \ k s^{-1}Mpc^{-1}
$$

\vskip 0.2in
\noindent{ \bf \it Type 1a SN \ (non-Key Project):}

One of the HST Key Project distance estimators involves the use of Type 
1a SN as standard candles.  As previously emphasized, the Key Project
does not perform direct measurements of Type 1a supernovae but rather
uses data obtained by other groups.   When these groups perform an
independent analysis to derive a value for the Hubble constant they
arrive at
 a smaller value than that quoted by the Key Project.  Their recent
quoted value is (\cite{1a}):

$$
H_0^{1a} = 64  ^{+8} _{-6} \ k s^{-1}Mpc^{-1}
$$

At the same time, Sandage and collaborators have performed an independent
analysis of SNe Ia distances and obtain (\cite{sandage}):

$$
H_0^{1a} = 58  \pm {6} \ k s^{-1}Mpc^{-1}
$$
\vskip 0.2in
\noindent{ \bf \it Surface Brightness Fluctuations and The Galaxy
Density Field:}

Another recently used distance estimator involves the measurement of
fluctuations in the galaxy surface brightness, which correspond to
density fluctuations allowing an estimate of the physical size of a
galaxy.  This measure yields a slightly higher value for the Hubble
constant (\cite{davis}):

$$
H_0^{SBF} = 74  \pm 4 \ k s^{-1}Mpc^{-1}
$$

\vskip 0.2in
\noindent{ \bf \it Time Delays in Gravitational Lensing:}

One of the most remarkable observations associated with observations of
multiple images of distant quasars due to gravitational lensing
intervening galaxies has been the measurement of the time delay in the
two images of quasar $Q0957 + 561$.  This time delay, measured quite
accurately to be $ 417 \pm 3$ days is due to two factors:  The
path-length difference between the quasar and the earth for the light
from the two different images, and the Shapiro gravitational time delay
for the light rays  traveling in slightly different
gravitational potential wells.  If it were not for this second factor, a
measurement of the time delay could  be directly used to determine the
distance of the intervening galaxy.  This latter factor however, 
implies that a model of both the galaxy, and the cluster in which it is
embedded must be used to estimate the Shapiro time delay.  This
introduces an additional model-dependent uncertainty into the analysis. 
Two different analyses yield values (\cite{chae}):

$$
H_0^{TD1} = 69  ^{+18} _{-12} (1-\kappa) \ k s^{-1}Mpc^{-1}
$$

$$
H_0^{TD2} = 74  ^{+18} _{-10} (1-\kappa) \ k s^{-1}Mpc^{-1}
$$
where $\kappa$ is a parameter which accounts for a possible deviation in 
 cluster parameters governing the overall induced gravitational time delay
of the two signals from that assumed in the best fit.  It is assumed in
the analysis that $\kappa$ is small.

\vskip 0.2in
\noindent{ \bf \it Summary:}

It is difficult to know how to best incorporate all of the quoted
estimates into a single estimate, given their separate systematic and
statistical uncertainties.  Assuming large number statistics, where large
here includes the quoted values presented here, I perform a simple weighted average
of the individual estimates, and find an approximate average value:

\begin{equation}
H_0^{Av} \approx 70  \pm 5 \ k s^{-1}Mpc^{-1}
\end{equation}

\subsubsection{Geometry:}

Again, for much of the 20th century the effort to determine the geometry of the Universe involved a very indirect route.  Einstein's Equations yield a relationship between the Hubble constant, the energy density, and the curvature of the Universe.   By attempting to determine the first two quantities, one hoped to constrain the third.  The problem is that until the past decade the uncertainty in the Hubble constant was at least 20-30 $ \%$ and the uncertainty in the average energy density of the universe was even greater.   As a result, almost any value for the net curvature of the universe remained viable.    

It has remained a dream of observational cosmologists to be able to
directly measure the geometry of space-time rather than infer the
curvature of the universe by comparing the expansion rate to the mean
mass density.   While several such tests, based on measuring galaxy counts
as a function of redshift, or the variation of angular diameter
distance with redshift, have been attempted in the past, these have all
been stymied by the achilles heel of many observational measurements in
cosmology, evolutionary effects.

Recently, however, measurements of the cosmic microwave background have
finally brought us to the threshold of a direct measurement of geometry,
independent of traditional astrophysical uncertainties.   The idea behind
this measurement is, in principle, quite simple.  As shown in
figure 2, the CMB originates from
a spherical shell located at the surface of last scattering (SLS), at a
redshift of roughly $z \approx 1000)$:

\begin{figure}
\centerline{\psfig{figure=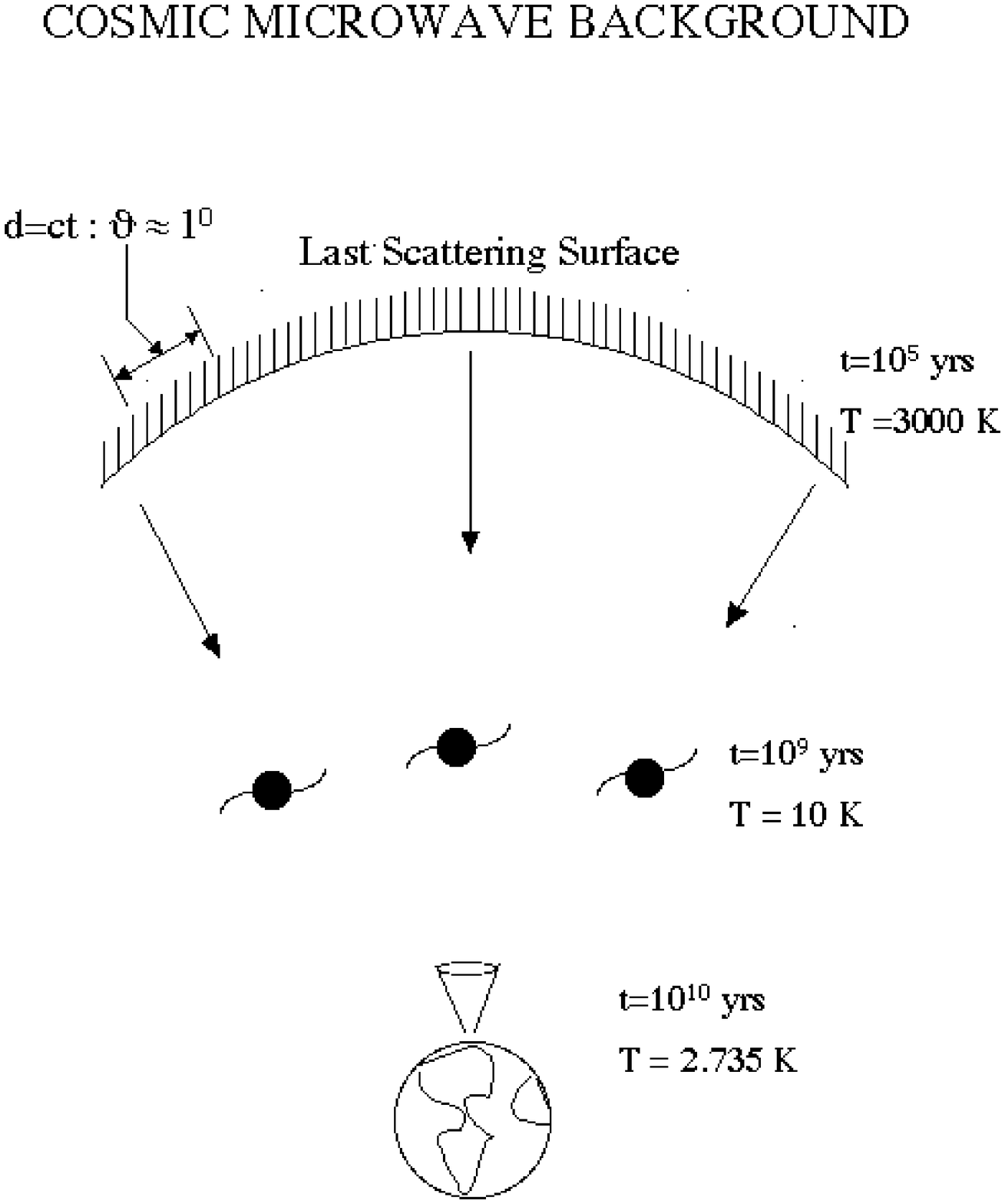,width=9.0cm}}
\label{cmb1}
\caption{A schematic diagram of the surface of last scattering, showing
the distance traversed by CMB radiation.}
\end{figure}

If a fiducial length could unambigously be distinguished on this surface,
then a determination of the angular size associated with this length
would allow a determination of the intervening geometry, as shown in figure 3.

\begin{figure}
\centerline{\psfig{figure=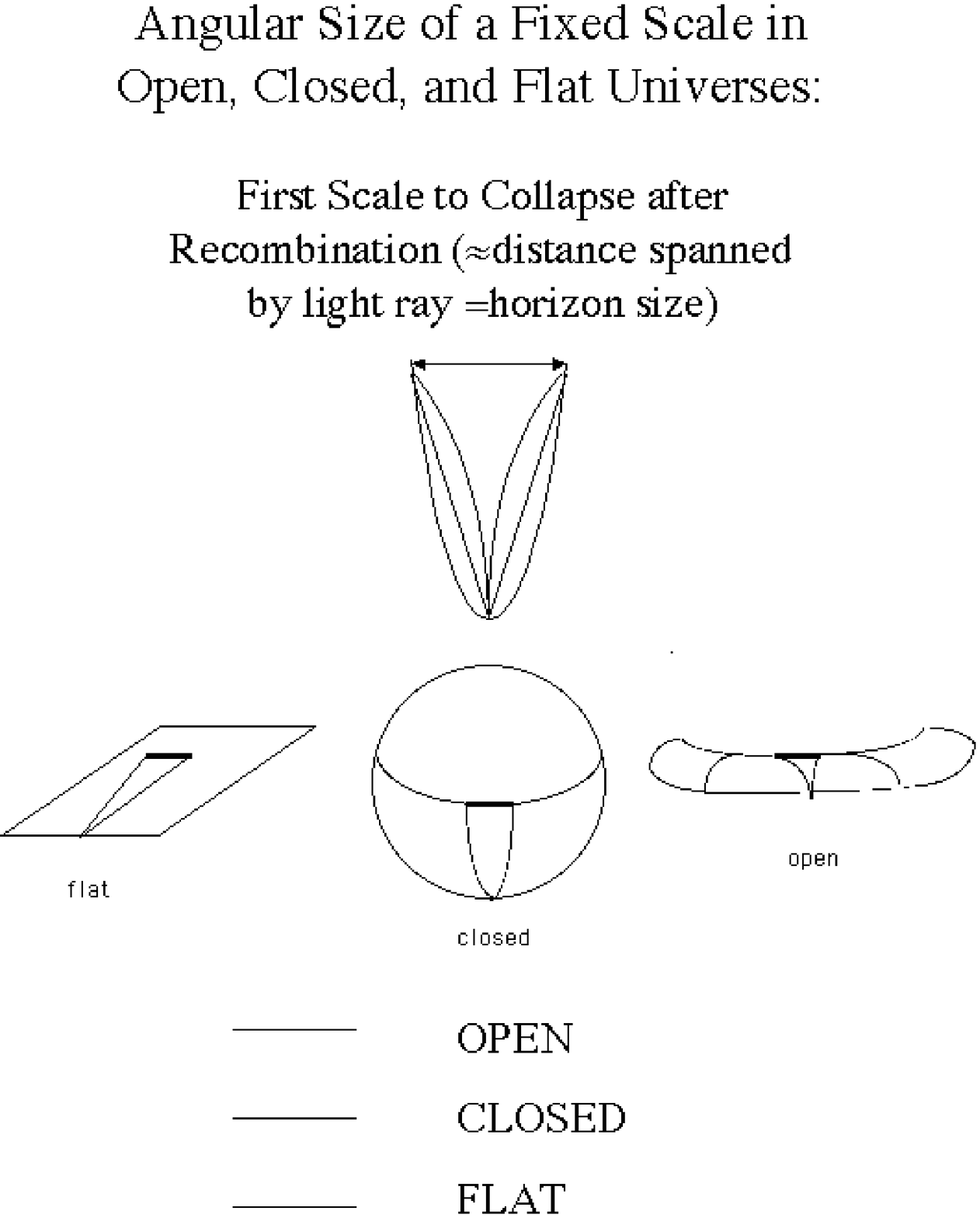,width=9.0cm}}
\label{cmb2}
\caption{The geometry of the Universe and ray trajectories for CMB
radiation.}
\end{figure}

Fortunately, nature has provided such a fiducial length, which
corresponds roughly to the horizon size at the time the surface of last
scattering existed (In this case the length is the "sound horizon", but since the medium
in question is relativistic, the speed of sound is close to the speed of light.)  
The reason for this is also straightforward.  This
is the largest scale over which causal effects at the time of the
creation of the surface of last scattering could have left an imprint. 
Density fluctuations on such scales would result in acoustic oscillations
of the matter-radiation fluid, and the doppler motion of electrons moving
along with this fluid which scatter on photons emerging from the SLS
produces a characteristic peak in the power spectrum of fluctuations of
the CMBR at a wavenumber corresponding to the angular scale spanned by
this physical scale.  These fluctuations should also be visually
distinguishable in an image map of the CMB, provided a resolution on
degree scales is possible.

A number of different ground-based balloon experiments, launched in
places such Texas and Antarctica have resulted in maps with the
required resolution (\cite{boomerang,maxima,vsa,dasi}).  Shown in figure 4 is a comparison
of the  Boomerang map with several simulations based on a gaussian
random spectrum of density fluctuations in a cold-dark matter universe,
for open, closed, and flat cosmologies.  Even at this qualitative level,
it is clear that a flat universe provides better agreement to between the
simulations and the data than either an open or closed universe.

\begin{figure}
\centerline{\psfig{figure=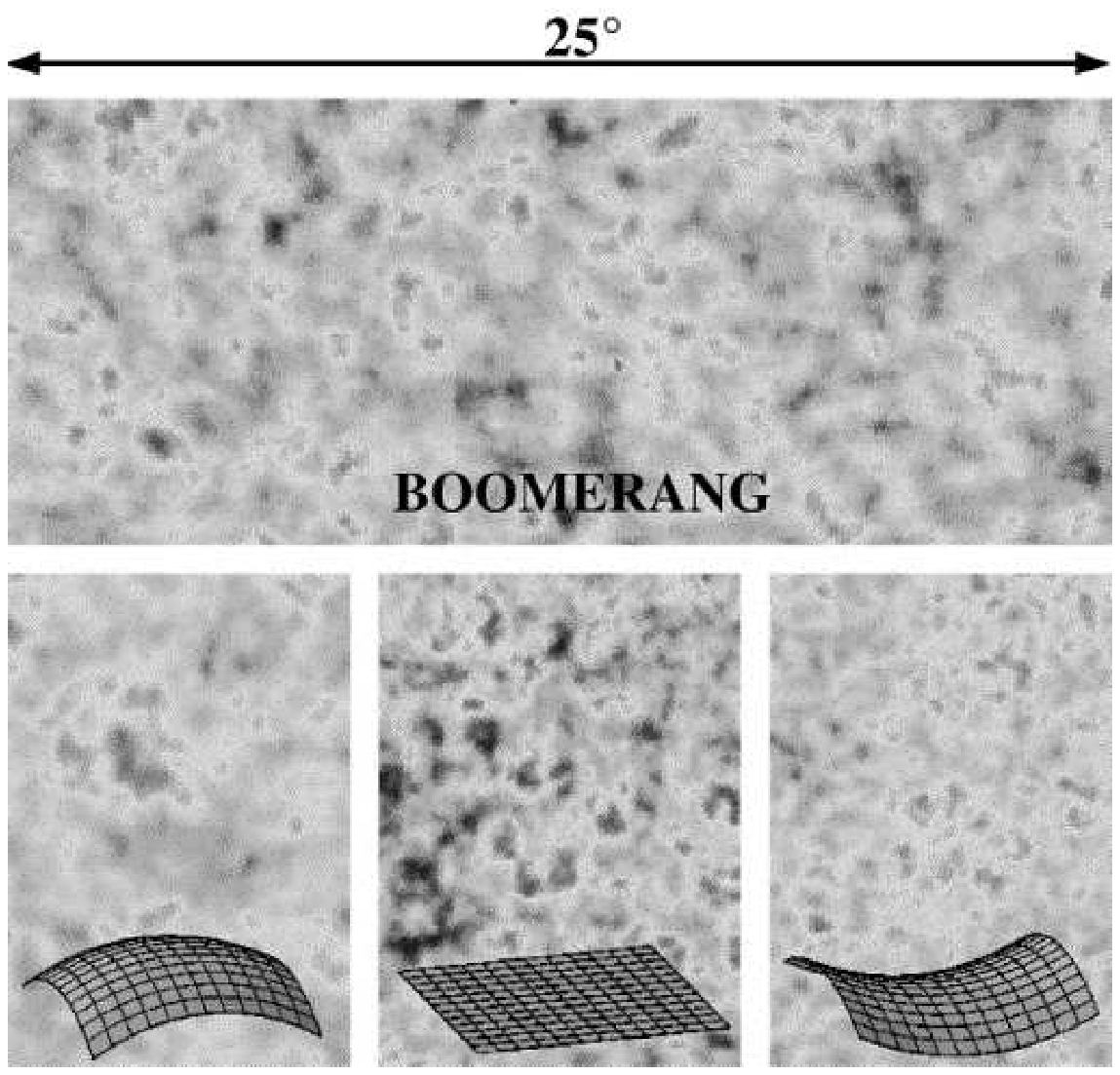,width=9.0cm}}
\label{cmb3}
\caption{Boomerang data visually compared to expectations for
an open, closed, and flat CDM Universe.}
\end{figure}

Recently the Wilkinson Microwave Anisotropy Probe (WMAP) has produced a high
resolution CMB map of the entire sky.   
Using this on can produce a quantitative constraint by comparing the inferred power spectra
with predicted spectra (\cite{teg}).
Such comparisions for the most recent data (\cite{WMAP}) yields a
constraint on the density parameter:

\begin{equation}
 \Omega = 1.02 \pm {0.02} (68 \% CL)
\end{equation}

For the first time, it appears that the longstanding prejudice of
theorists, namely that we live in a flat universe, may have been
vindicated by observation!   However, theorists can not be too
self-satisfied by this result, because the source of this energy density
appears to be completely unexpected, and largely inexplicable at the
present time, as we will shortly see.

\subsection{Time}

\subsubsection{Stellar Ages:}

Ever since Kelvin and Helmholtz first estimated the age of the Sun to be
less than 100 million years, assuming that gravitational contraction was
its prime energy source, there has been a tension between stellar age
estimates and estimates of the age of the universe.   In the case of the
Kelvin-Helmholtz case, the age of the sun appeared too short to
accomodate an Earth which was several billion years old.  Over much of
the latter half of the 20th century, the opposite problem dominated the
cosmological landscape.    Stellar ages, based on nuclear reactions as
measured in the laboratory, appeared to be too old to accomodate even an
open universe, based on estimates of the Hubble parameter.  Again, as I
shall outline in the next section, the observed expansion rate gives an
upper limit on the age of the Universe which depends, to some degree, 
upon the equation of
state, and the overall energy density of the dominant matter in the
Universe.  

There are several methods to attempt to determine stellar ages, but I
will concentrate here on main sequence fitting techiniques, because those
are the ones I have been involved in.  For a more general review, see \cite{krausschabsci}.

The basic idea behind main sequence fitting is simple.  A stellar model is
constructed by solving the basic equations of stellar structure, including
conservation of mass and energy and the assumption of hydrostatic equilibrium,
and the equations of energy transport.  Boundary conditions at the center
of the star and at the surface are then used, and combined with assumed
equation of state equations, opacities, and nuclear reaction rates in
order to evolve a star of given mass, and elemental composition.

Globular clusters are compact stellar systems containing up to $10^5$ stars,
with low heavy element abundance.  Many are located in a spherical halo around
the galactic center, suggesting they formed early in the history of our
galaxy.  By making a cut on those clusters with large halo velocities, and
lowest metallicities (less than 1/100th the solar value), one attempts to
observationally distinguish the oldest such systems. Because these systems are
compact, one can safely assume that all the stars within them formed at
approximately the same time.

Observers measure the color and luminosity of stars in such clusters, producing
color-magnitude diagrams of the type shown in figure 5 (based on data 
from \cite{durr}).

\begin{figure}
\centerline{\psfig{figure=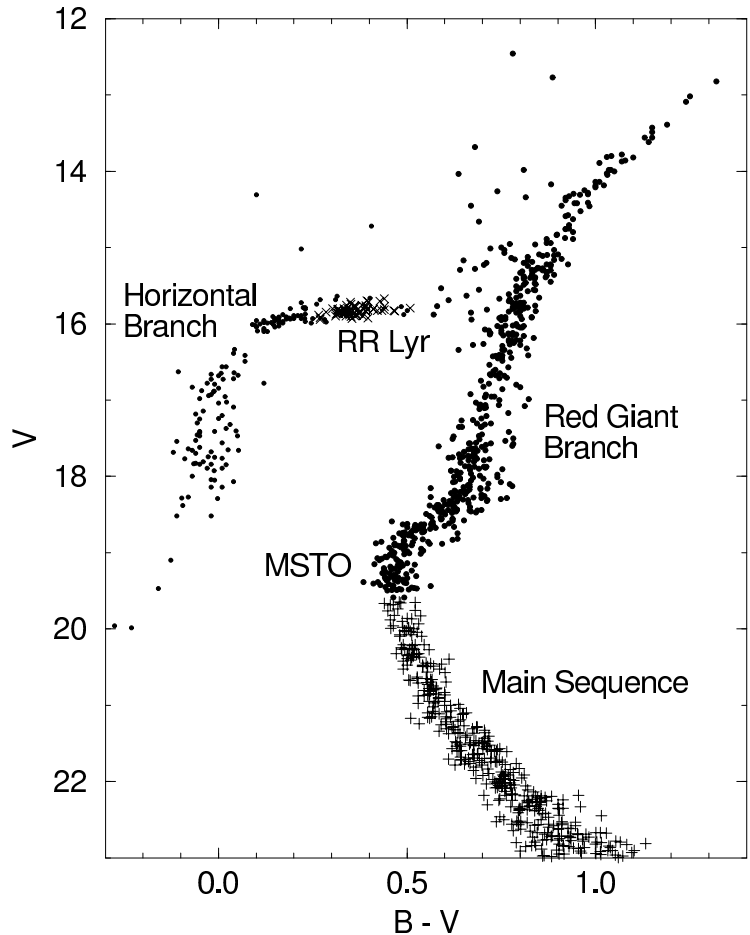,width=9.0cm}}
\label{gccmd}
\caption{Color-magnitude diagram for a typical globular
cluster, M15. Vertical axis plots the magnitude (luminosity) of
the stars in the V wavelength region and the horizontal axis plots the
color (surface temperature) of the stars.}
\end{figure}

Next, using stellar models, one can attempt to evolve stars of differing
mass for the metallicities appropriate to a given cluster, in order to fit
observations.  A point which is often conveniently chosen is the so-called main
sequence-turnoff (MSTO) point, the point in which hydrogen burning (main
sequence) stars have exhausted their supply of hydrogen in the core.  After the
MSTO, the stars quickly expand, become brighter, and are referred to as Red
Giant Branch (RGB) stars.  Higher mass stars develop a helium core that is so
hot and dense that helium fusion begins.  These form along the horizontal
branch.  Some stars along this branch are unstable to radial pulsations, the
so-called RR Lyrae stars mentioned earlier, which are important distance
indicators.  While one in principle could attempt to fit theoretical isochrones
(the locus of points on the predicted CM curve corresponding to different mass
stars which have evolved to a specified age), to observations at any point, the
main sequence turnoff is both sensitive to age, and involves minimal
(though just how minimal remains to be seen) theoretical uncertainties.

Dimensional analysis tells us that the main sequence turnoff should be a
sensitive function of age.  The luminosity of upper main sequence stars is very
roughly proportional to the third power of solar mass.  Hence the time it takes
to burn the hydrogen fuel is proportional to the total amount of fuel
(proportional to the mass M), divided by the Luminosity---
proportional to $M^3$.  Hence the lifetime of stars on the main sequence
is roughly proportional to the inverse square of the stellar mass.

Of course the ability to go beyond this rough approximation depends completely
on the on the confidence one has in one's stellar models.  What is most important for
the comparison of cosmological predictions with inferred age estimates is
the uncertainties in stellar model parameters, and
not merely their best fit values. 

Over the course of the past several years, I and my collaborators have
tried to incorporate stellar model uncertainties, along with
observational uncertainties into a self consistent Monte Carlo analysis
which might allow one to estimate a reliable range of globular cluster
ages.   Others have carried out independent, but similar studies, and at
the present time, rough agreement has been obtained between the different
groups (i.e. see\cite{krauss}).  

I will not belabor the detailed history of all such efforts here.  The
most crucial insight has been that stellar model uncertainties
are small in comparison to an overall observational uncertainty inherent
in fitting predicted main sequence luminosities to observed turnoff
magnitudes.  This matching depends crucially on a determination of the
distance to globular clusters.  The uncertainty in this distance scale
produces by far the largest uncertainty in the quoted age estimates.

In many studies, the distance to globular clusters can be parametrized in
terms of the inferred magnitude of the horizontal branch stars.  This
magnitude can, in turn, be presented in terms of the inferred absolute
magnitude,
$M_v(\rm RR)$of RR Lyrae
variable stars located on the horizontal branch. 

In 1997, the Hipparcos satellite produced its catalogue of 
parallaxes of nearby stars,
causing an apparent revision in distance estimates.  The Hipparcos parallaxes
seemed to be systematically smaller, for the smallest measured parallaxes, than
previous terrestrially determined parallaxes.  Could this represent the
unanticipated systematic uncertainty that David has suspected?  Since all 
the
detailed analyses had been pre-Hipparcos, several groups scrambled to
incorporate the Hipparcos catalogue into their analyses.  The immediate result
was a generally lower mean age estimate, reducing the mean value to 11.5-12 Gyr,
and allowing ages of the oldest globular clusters as low as 9.5 Gyr.   However,
what is also clear is that there is now an explicit systematic uncertainty in
the RR Lyrae distance modulus which dominates the results.  Different
measurements are no longer consistent.  Depending upon which distance estimator
is correct, and there is now better evidence that the distance estimators which
disagree with Hipparcos-based main sequence fitting should not be dismissed out
of hand, the best-fit globular cluster estimate could shift up perhaps $1
\sigma$, or about 1.5 Gyr, to about 13 Gyr.

Within the past two years, Brian Chaboyer and I have reanalyzed globular
cluster ages, incorporating new nuclear reaction rates, cosmological
estimates of the $^4$He abundance, and most importantly, several new
estimates of $M_v(\rm RR)$, shown below.   

\begin{figure}
\centerline{\psfig{figure=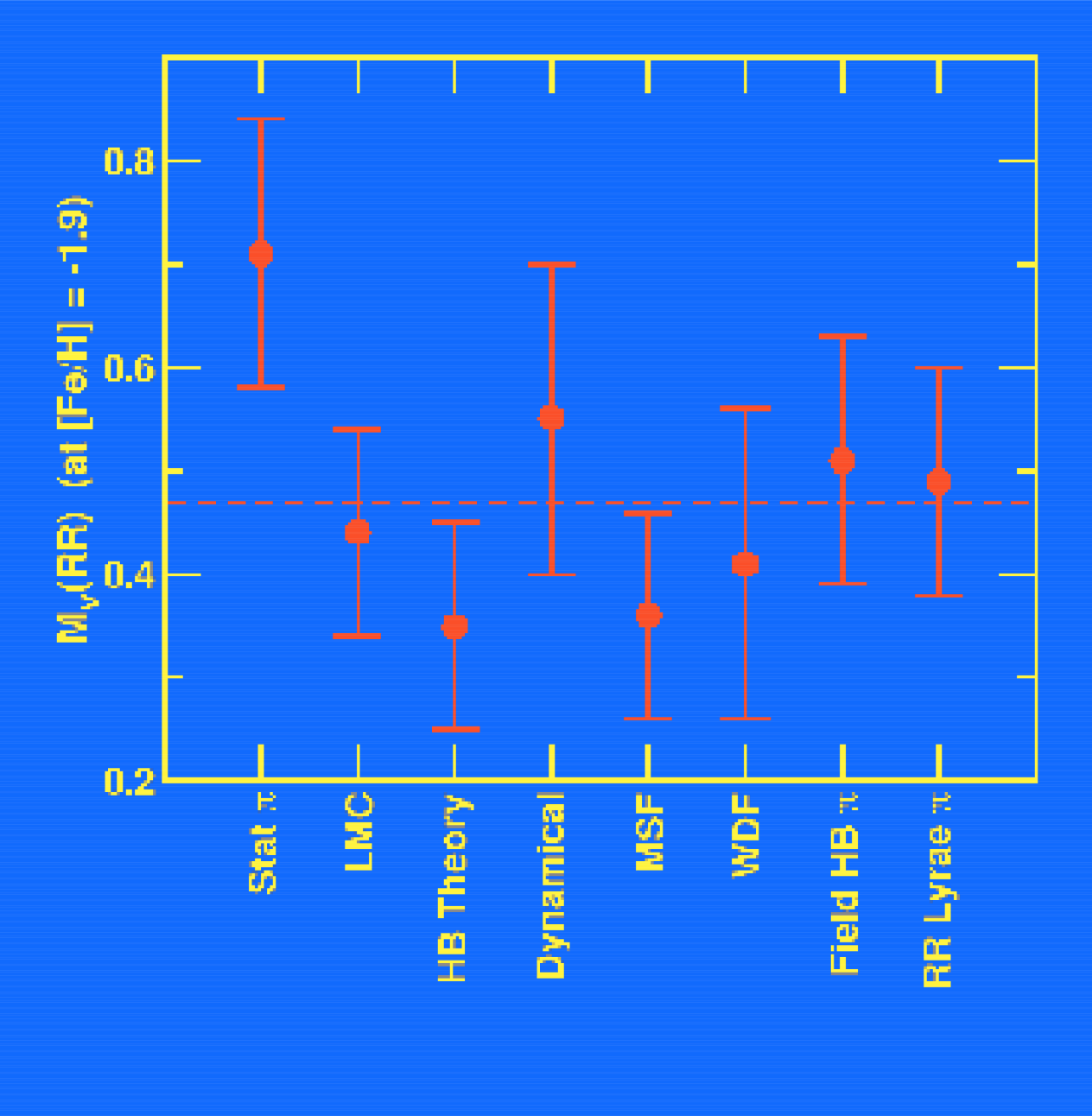,width=9.0cm}}
\label{mvrr}
\caption{Different estimates of the inferred magnitude of horizontal branch RR Lyrae 
stars, with uncertainties}
\end{figure}

The result is that while systematic
uncertainties clearly still dominate, we argue that the best fit age of globular clusters
is now
$12.6^{+3.4}_{-2.4} $ ($95\%$) Gyr, with a 95 $\%$ confidence range of about
11-16 Gyr (\cite{krausschabsci}).   

\begin{figure}
\centerline{\psfig{figure=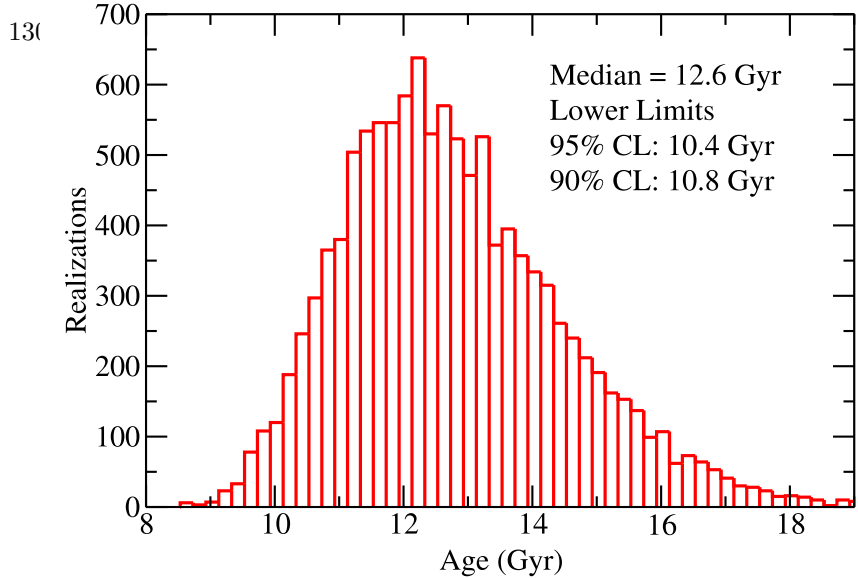,width=9.0cm}}
\label{hist}
\vspace{2pc}
\caption{Histogram showing range of age fits to old globular clusters using Monte
Carlo analysis}
\end{figure}

If we are to turn this result into a lower limit on the age of the Universe we must add to this estimate the time after the Big Bang that it took for the first globular clusters in our galaxy to form.   Here there is great uncertainty.  However a robust lower limit comes from observations of structure formation in the Universe, which suggest that the first galaxies could not have formed much before a redshift of 6-7.   Turning this redshift into an age depends upon the equation of state of the dominant energy density at that time (see below).  However, one can show that at such high redshifts, the effects of a possible dark energy component are minimal, leading to a minimum age of globular cluster formation of about .8 Gyr.   The maximum age is much less certain, as it is possible for galaxies to form at redshifts as low as 1-2.  Thus, one must add an age of perhaps 3.5-4 Gyr to the globular age estimate above to get an upper limit on the age of the Universe.  Putting these factors together, one derives a 95$\%$ confidence age range for the Universe of 11.2-20 Gyr.

\subsubsection{Hubble Age:}

As alluded to earlier, in a Friedman-Robertson-Walker Universe, the age of
the Universe is directly related to both the overall density of energy,
and to the equation of state of the dominant component of this energy
density.  The equation of state is parameterized by the ratio $\omega = p
/{\rho}$, where $p$ stands for pressure and $\rho$ for energy
density.  It is this ratio which enters into the second order Friedman
equation describing the change in Hubble parameter with time, which in
turn determines the age of the Universe for a specific net total energy
density.

 The fact that this depends on two independent parameters has
meant that one could reconcile possible conflicts with globular cluster
age estimates by altering either the energy density, or the equation of
state.   An open universe, for example, is older for a given Hubble
Constant, than is a flat universe, while a flat universe dominated by a
cosmological constant can be older than an open matter dominated
universe.

If, however, we incorporate the recent geometric determination which
suggests we live in a flat Universe into our analysis, then our
constraints on the possible equation of state on the dominant energy 
density of the universe become more severe.   If, for existence, we
allow for a diffuse component to the total energy density with the
equation of state of a cosmological constant ($\omega =-1$), then the
age of the Universe for various combinations of matter and cosmological
constant  is given by:

\begin{equation}
H_0t_0 =  \int _{0}^{\infty }{dz \over{(1+z)  [(\Omega_{m})(1+z)^3  +  (\Omega_X)(1+z)^{3(1+w)}]^{1/2}}}
\end{equation}

This leads to ages as shown in Table 1.

\begin{table}
\caption{Hubble Ages for a Flat Universe, $H_0 = 70 \pm 8$, }
\begin{center}
\begin{tabular}{|l|l| c|}
\hline
 $\Omega_M$ & $\Omega_x$ & $t_0$ \\ \hline
 $1$ & $0$ & $9.7 \pm 1$ \\
 $0.2$ & $0.8$ & $15.3 \pm 1.5$ \\
 $0.3$ & $0.7$ & $13.7 \pm 1.4$ \\
 $0.35$ & $0.65$ & $12.9 \pm 1.3$ \\ \hline
\end{tabular}
\end{center}
\end{table}

The existing limits on the age of the universe from globular clusters are thus already
are {\it incompatible} with a flat matter dominated universe.   This is a very important result, as it implies that now all three classic tests of cosmology, including geometry, large scale structure, and age of the Universe now support the same cosmological model, which involves a universe dominated by dark energy  (Indeed, before the direct supernova evidence for dark energy it was argued that these factors favored the existence of dark energy (\cite{kraussturn1}) .  We can provide limits on
the equation of state for dark energy as well.  Shown in figure 8, is the constraint on $w$,
assuming a Hubble constant of 72  (\cite{krausschabsci}).

\begin{figure}
\vspace{2pc}
\centerline{\psfig{figure=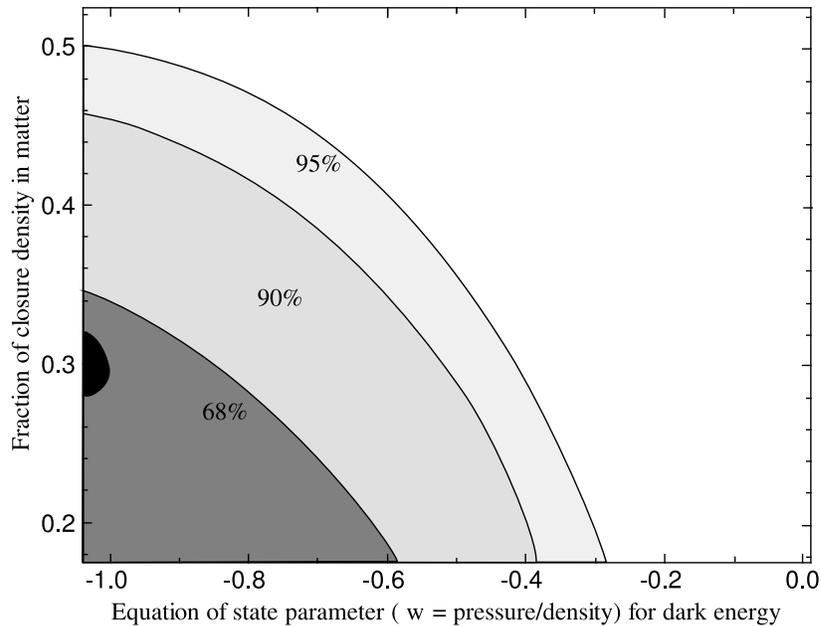,width=7.0cm}}
\label{darke}
\vspace{2pc}
\caption{Constraint on the equation of state parameter for dark energy as a function of
the fraction of closure density in matter resulting from age constraint described here.}
\label{fig:mvrr}
\end{figure}

At the same time, it is worth noting that unfortunately the upper limit on the age of the universe coming from globular cluster ages cannot provide a useful limit on the equation of state parameter $w$, because there is an upper limit on the Hubble Age, independent of $w$, if the contribution of matter to the total density is greater than 20$\%$ (\cite{krauss03}).

\subsubsection{CMB, Hubble Age, Galaxy Formation and Equation of State}

Perhaps not so remarkably, the CMB can also give a direct measure of the Hubble Age, in much the
same way as one can use it to measure the geometry of the Universe.  The physical distance to the
last scattering surface depends upon the age of the Universe, so that measuring the physical angle of the first doppler peak can, provided one uses other observational information about $H_0$ and $\Omega_{matter}$, give a measure of the Hubble age.   The first such estimates were obtained with
Boomerang data, but once again the WMAP data gives the best current limit, which is quoted
as (\cite{WMAP})  $13.7 \pm 0.2$ Gyr.   

By comparing WMAP observations with previous estimates of globular cluster ages, one can derive provide important new handles to probe the likely formation of the milky way galaxy, and in a broader sense the formation of large scale cosmic structures.  The two key WMAP observations in this regard are the estimate of cosmic age ($ 13.7 \pm 0.2$ Gyr), and the redshift of reionization (\cite{WMAP}).  

Comparing the 68$\%$ lower confidence limit age of 11.2 Gyr (\cite{krausschabsci}) with the 68$\%$ upper limit on the age of the Universe from WMAP of $13.9$ Gyr suggests an $90\%$ upper limit  $\approx 2.7$ Gyr as the time after the Big Bang that globular clusters in our galaxy first formed from the primordial halo of gas that ultimately collapsed to form the Milky Way.  At the $ 95\%$ confidence level the limit becomes approximately 3 Gyr.   This not only improves upon previous estimates, it is the first direct constraint on this quantity.

Of somewhat more interest is a determination of the most probable time after the Big Bang at which our globular clusters formed.  Now that WMAP has determined a surprisingly early time where the Universe reionized, corresponding to an age of about 200-300 Myr after the Big Bang, it is interesting to know whether this corresponds to an early period of star formation, and whether structures as large as globular clusters of stars also formed this early.  Note that (\cite{jimenez}) have recently assumed this to be the case. 

A variety of different methods have been used to determine the age of globular clusters in our galaxy.  The Monte Carlo analysis referred to above involves dating these clusters using main-sequence turnoff luminosity, and yields an age estimate for the oldest clusters at the 95$\%$ confidence level of $12.6\pm ^{3.4}_{2.2}$ Gyr.  The most likely age for these globular clusters is thus $\approx$ 800 Myr younger than the WMAP lower limit on the age of the Universe.   However, because the distribution is broad, the possibility that globular clusters formed before the period of cosmic reionization determined by WMAP to occur $\approx$ 200 Myr after the Big Bang is certainly still viable (\cite{jimenez}).    Nevertheless, examining the probability distribution in \cite{krausschabsci}, as fit analytically in \cite{jimenez}, one finds a 75$\%$ likelihood that the oldest globular clusters are in fact less than 13.5 Gyr old.

While not compelling, the possibility that globular clusters in our galaxy may have formed well after reionization could shed light on a number of issues, including whether reionization is due to a very early generation of massive stars and whether such systems formed  before (and if so, how much before) larger structures such as globular clusters.  This could probe the nature of possible hierarchical clustering.   The likelihood of this possibility is increased when one recognizes that several other methods for determining the age of globular clusters, including using luminosity functions (\cite{jimenez2}), white dwarf cooling (\cite{hansen}) and eclipsing binaries (\cite{krausschab2}) favor globular cluster ages in the range of 11-13 Gyr.   

The existing uncertainty in globular cluster dating techniques is at present too large to do more than hint that there may be a gap in time between  reionization in the Universe and the formation of larger scale structures.  However, this hint  strongly motivates efforts to further reduce the absolute uncertainty in globular dating techniques.  

Next, one can  use the Hubble Age determination from the CMB to constrain the possibility that the
equation of state parameter for dark energy is actually less than $-1$.  While there are really no sensible models of this, there is also no understanding  of the dark energy, so who knows?

Age estimates can in principle give strong constraints on values of $w$ less than -1, since  the age of the Universe is a strongly varying function of $w$ for values of $w >-5$ (\cite{krauss03}).

In Figs.9 and 10, the predicted age of the Universe for various values of $w < -1$ as a function of the Hubble constant in comparison to the $2\sigma$ upper limit on the cosmic age from WMAP, for two different values of the assumed matter density today(corresponding to midpoint of  the WMAP allowed range for matter density, and the $2\sigma$ upper limit)(\cite{krauss03a}).   As is clear from these figures, for a flat universe the inferred bound on $w$ from the WMAP cosmic age limit depends sensitively on the assumed total matter density today.  

It is important to realize however that one is not free to independently vary $\Omega_m$ and $H_0$ in deriving bounds using the WMAP data.   These two quantities are themselves highly anti-correlated in the WMAP fit (\cite{WMAP}.   As can be seen from the WMAP fits  as $w$ is decreased (for $w>-1$), the allowed range of $\Omega_m$ decreases roughly linearly, while the allowed range of Hubble constant increases roughly linearly.  If we assume this behavior extrapolates to values of $w <-1$, then we can use this relation to derive a conservative lower bound on $w$.   The most conservative bound on $w$ comes from assuming the largest allowed value of $\Omega_m$ for any value of $H_0$.   We fit this value using the anti-correlation described above, and fitting to the WMAP plots to derive $\Omega_m^{max} h^2 =0.309-0.243h$ within the allowed range of $H_0$.   When we include this relation explicitly for $\Omega_m$ in the cosmic age relation, one derive limits on the age of the Universe shown in Fig. 11 (\cite{krauss03a}). 

To derive a bound on $w$ it is necessary to note that lower bound on $H_0$ derived from WMAP is correlated with the inferred value of $w$.   If we extrapolate the allowed range of $H_0$ to values of $w < -1$, we find a lower bound on $H_0$ as a function of $w$  shown by the thick solid line in this figure.   If we use this lower bound on $H_0$, and compare the predicted age as a function of $w$ with the WMAP upper limit, we derive a bound $w>-1.22$.   If we were instead to allow the full HST range for $H_0$ in deriving this limit, the lower bound would decrease slightly to $w>-1.27$.

The WMAP team has used a global fit to constrain $w$ and they find the lower bound $w >-1.2$.  If one combines this result with the WMAP-derived upper bound on $w$, one thus finds an allowed region $-1.2<w< -.8$.    It is interesting, but perhaps not surprising that the uncertainty is symmetric about the value -1.  I shall have more to say about this later.

\begin{figure}
\centerline{\psfig{figure=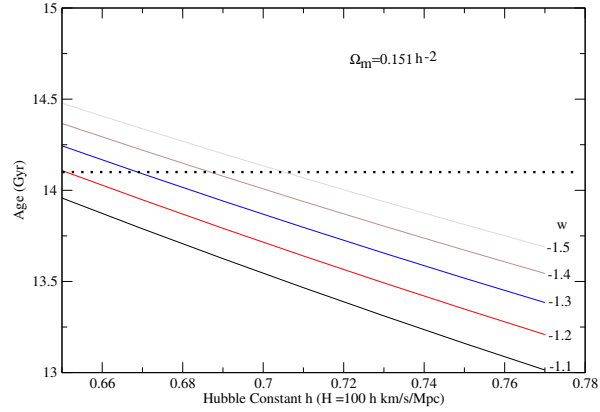,width=9.0cm}}
\label{f1}
\caption{Shown are contours of cosmic age versus Hubble constant for various constant values of the equation of state parameter for dark energy, $w <-1$, and for the matter density taking its maximum value within the $2 \sigma$ range given by WMAP, i.e. $\Omega_{m} h^2 =0.151$.  Also shown (dotted line) is the WMAP upper cosmic age constraint, where it is assumed that the age limit is not correlated with the values of  the Hubble constant within the range allowed by WMAP. }
\end{figure}

\begin{figure}
\centerline{\psfig{figure=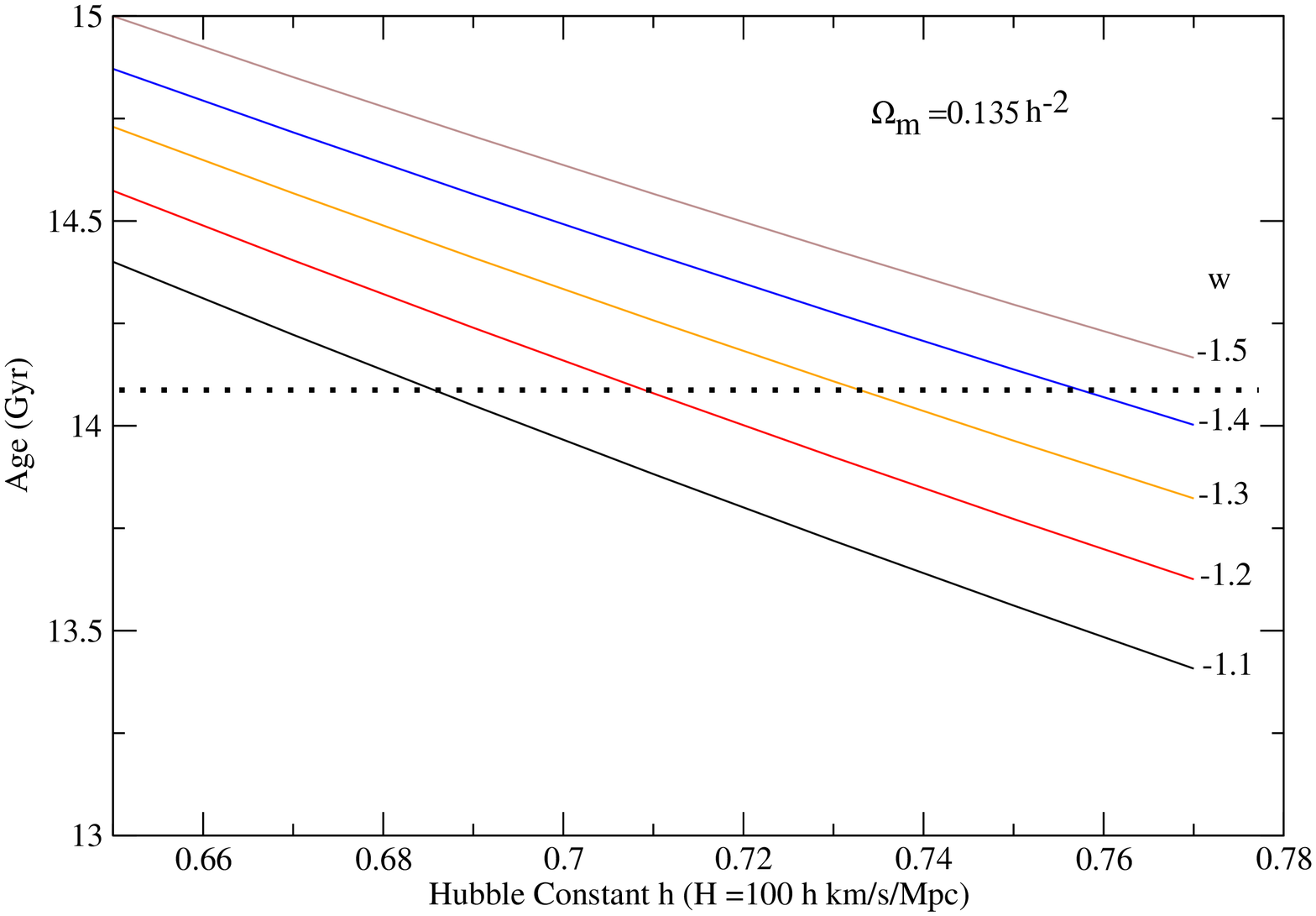,width=9.0cm}}
\label{f2}
\caption{Same as Figure 9, for the midpoint value  $\Omega_{m} h^2 =0.135$ within the $2 \sigma$ range given by WMAP}.   
\end{figure}

\begin{figure}
\centerline{\psfig{figure=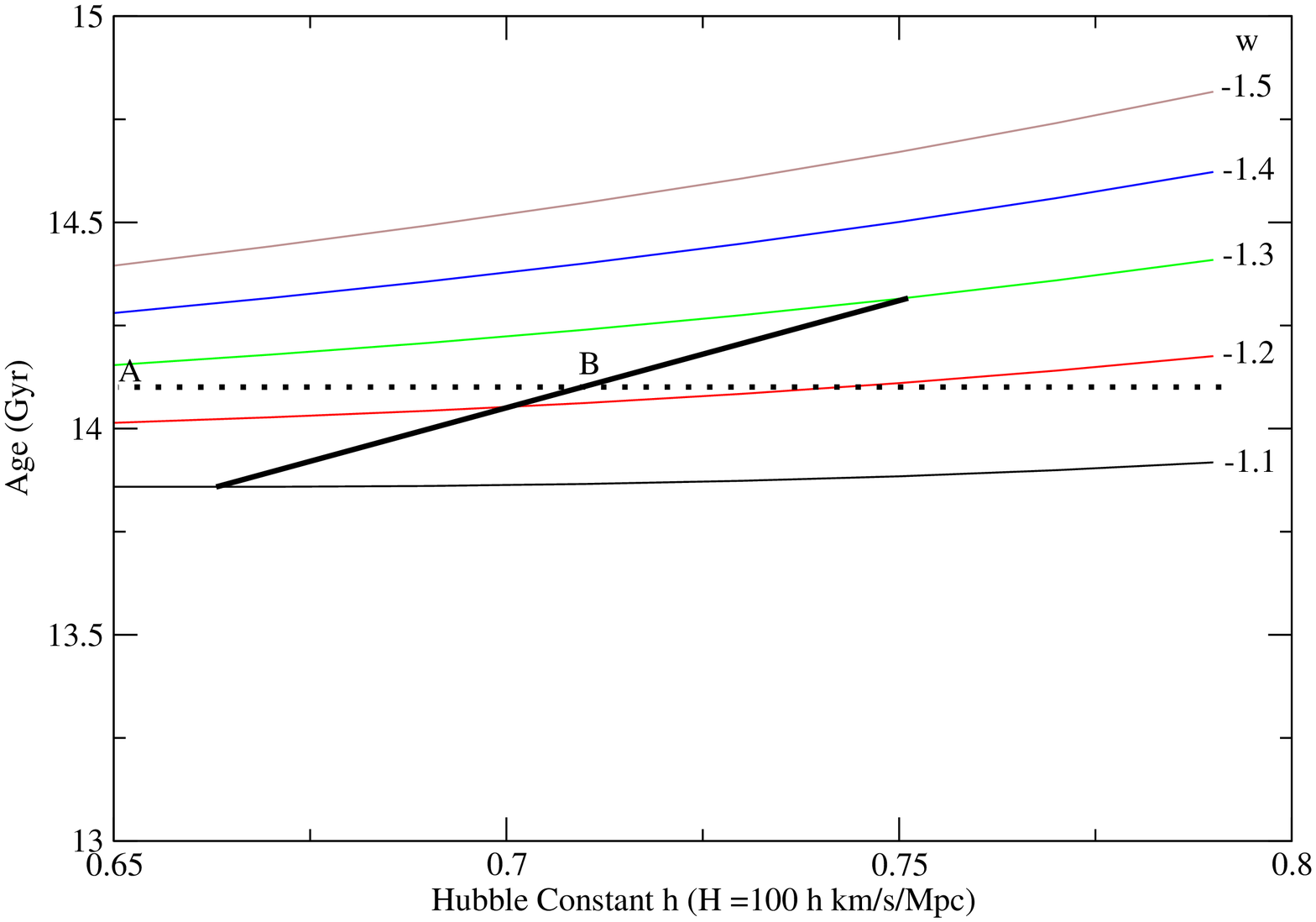,width=9.0cm}}
\label{f3}
\caption{Assuming a negative linear correlation between inferred value of the density parameter 
$\Omega_{m} h^2$ and the inferred value of $h$, based on the WMAP data, age estimates as a function of equation of state parameter $w$ can be determined.  Because predicted age is a decreasing function of the density parameter, the most conservative limits on $w$ come from choosing the maximum allowed density parameter (at the $2\sigma$ level) from WMAP. A fit to WMAP yields ($\Omega_{m}^{max}h^2=0.309-.243h$). Also shown (dotted line) is the WMAP upper cosmic age constraint, where it is assumed that the age limit is not correlated with the values of  the Hubble constant within the range allowed by WMAP.   Finally, also shown (solid curve) is the inferred lower bound on $h$ as a function of $w$ estimated by extrapolating WMAP plots.   The constraints on $w$ derived with and without this extra constraint are obtained at the points B, and A respectively.  }
\end{figure}

\subsection{Matter}

It has long been established that visible matter, namely matter associated with stars, planets,
and luminous gas falls far short of the amount required to close the Universe.   The best
estimate today of the total visible matter density yields $\Omega_{lum} = .004 exp(\pm 0.3) $. 

The problem is that this estimate is clearly a lower limit on the total baryon density in the universe, since matter need not shine.    In order to directly determine the total baryon abundance we have been
able to resort to calculations from the early universe of the light element abundances, and then comparing these to observations, as I describe below

\subsubsection{The Baryon Density: a re-occuring crisis?:}

The success of Big Bang Nucleosynthesis in predicting in the cosmic
abundances of the light elements has been much heralded.  The basis
  is quite simple: At $T = 10^9 - 10^{10} K$ nuclear reactions convert protons and neutrons
  to $^4He$ via intermediate reactions that produce $D$ and $^3He$.   Since reaction rates
  depend upon the density of protons and neutrons, this ultimately depends upon $\Omega_B$.
  The final important fact is that production of $^4He$ cannot begin until sufficient $D$ is
  produced so that further reactions processing $D$ to $He$ can take place.  
  
  The greater the density of protons and neutrons, the more efficiently, and the earlier $^4He$ is 
  produced.  Thus the remnant $^4He$ abundance is a monotonically increasing function 
  of $\Omega_B$.   Similarly the more efficiently $^4He$ is produced, the more efficiently
  $D$ and $^3He$ are burned to produce it, and thus these remnant abundances are monotonically
  decreasing functions of the baryon density.

 The finer the ability to empirically infer the primordial abundances on
the basis of observations, the greater the ability to uncover some
small deviation from the predictions.   Over the past five years, two
different sets of observations have threatened, at least in some people's
minds, to overturn the simplest BBN model predictions.  I believe it is
fair to say that most people have accepted that the first threat was
overblown.  The concerns about the second have only recently subsided.
\vskip 0.1in

{\it i. Primordial Deuterium:}  As noted above, the production of primordial deuterium
during BBN is a monotonically decreasing function of the baryon density
simply because the greater this density the more efficiently protons and
neutrons get processed to helium, and deuterium, as an intermediary
in this reactions set, is thus also more efficiently processed at the
same time.   The problem with inferring the primordial deuterium
abundance by using present day measurements of deuterium abundances in
the solar system, for example, is that deuterium is highly processed
(i.e. destroyed) in stars, and no one has a good enough model for
galactic chemical evolution to work backwards from the observed
abundances in order to adequately constrain deuterium at a level where
this constraint could significantly test BBN estimates.

Five years ago, the situation regarding deuterium as a probe of BBN
changed dramatically, when David Tytler and Scott Burles convincingly
measured the deuterium fraction in high redshift hydrogen clouds that
absorb light from even higher redshift quasars.   Because these clouds
are at high redshift, before significant star formation has occurred,
little post BBN deuterium processing is thought to have taken place, and
thus the measured value gives a reasonable handle on the primordial BBN
abundance.  The best measured system (\cite{tytler}) yields a deuterium to
hydrogen fraction of 

\begin{equation}
(D/H) = (3.3. \pm 0.8) \times 10^{-5} \ \  ( 2 \sigma )
\end{equation}

This, in turn, leads to a contraint on the baryon fraction of the
Universe, via standard BBN, 

\begin{equation}
\Omega_{B} h^2 = .02 \pm .002 \ \  ( 2 \sigma )
\end{equation}

where the quoted uncertainty is dominated by the observational
uncertainty in the D/H ratio, and where $H_0 =100 h$.  Thus, taken at face
value, we now know the baryon density in the universe today to an
accuracy of about
$10 \%$!

When first quoted, this result sent shock waves through some of the
BBN community, because this value of $\Omega_B$ is only consistent if the
primordial helium fraction (by mass) is greater than about 24.5$\%$. 
However, a number of previous studies had claimed an upper limit well
below this value.  However, recent studies, for example, place an upper limit
on the primordial helium fraction closer to 25$\%$.

\vskip 0.1in
{\it ii. CMB constraints:} Beyond the great excitement over the
observation of a peak in the CMB power spectrum at an angular scale
corresponding to that expected for a flat universe lay some
excitement/concern over the small apparent size of the next peak in the
spectrum, at higher multipole moment (smaller angular size).  The height
of the first peak in the CMB spectrum is related to a number of
cosmological parameters and thus cannot alone be used to constrain any
one of them.  However, the relative height of the first and second peaks
is strongly dependent on the baryon fraction of the universe, since the
peaks themselves arise from compton scattering of photons off of
electrons in the process of becoming bound to baryons.   Analyses of the
two first small-scale CMB results originall produced a constraint which was in disagreement with the BBN estimate.  However, more recent data indicates $\Omega_Bh^2=0.021$,  precisely where one would expect it to be based on BBN predictions.
  
Most recently reported measurements of $^3He$ in the Milky Way Galaxy give the constraint,  
$^3He/H =(1.1. \pm 0.2) \times 10^{-5}$, which in turn implies
 $\Omega_Bh^2=0.02$.  Thus, all data is now consistent with  the assumption that the Burles and
Tytler limit on
$\Omega_B h^2$ is correct, adding further confidence in the predictions of BBN.  Taking the range for $H_0$ given earlier,
one derives the constraint on
$\Omega_B$ of

\begin{equation}
\Omega_{B} = .045 \pm 0.15
\end{equation}
 
 Note that the measured baryon density  is a factor of 5-10 greater than the measured density of luminous matter in the Universe.   Clearly, most baryons are dark.   The next question is, can these
 dark baryons account for all gravitating matter in the Universe?

\subsubsection{ $\Omega_{matter}$}
 
If one is interested in measuring gravitating matter, the best way to measure it is using gravity.
Indeed, one can measure the mass of the Sun by using Newton's relation for the velocity of 
objects in roughly circular orbits, $v^2 = GM/r$.   In the 1970's, this technique was used to 
attempt to measure the mass of our galaxy.  Our Sun orbits around the outer edge of
 our galaxy with a velocity of approximately 220 km/s, at a radius of approximately 8 kpc,
 implying a mass inside its orbit of approximately $10^11$ solar masses, in good agreement
 with mass estimates based on the total number of stars in our galaxy.   However, when 
 test objects, including globular clusters, gas clouds, and satellite galaxies that orbit at distances
 of up to 10 times the distance from the galactic center, far outside the luminous region, instead
 of falling off, velocities remain roughly constant.   Unless gravity is changing, this implies a 
 total mass that increases with radius from the center of the galaxy, thus implying at least
 90 $\%$ of the mass of our galaxy is dark.   What is more remarkable is that a similar behavior
 is observed in almost all spiral galaxies.  
 
 A mass that grows linearly would derive from a density distribution (assuming sphericity) that
 falls like $1/r^2$.    Interestingly enough, if one assumes a collisionless gas with isotropic
 initial velocity distribution  $<v^2> \approx $ constant, then its equation of state is given
 by 
 
 \begin{equation}
 p(r) = \rho (r) \sigma^2 = \rho (r) <(v_x - \bar{v}_x)^2>
 \end{equation}
 
 Then if one imposes the condition of hydrostatic equilibrium on the system, with pressure balancing
 gravity,
 
 \begin{equation}
 -{dp \over dr}  = {GM(r) \over r^2}  \rho (r)
 \end{equation}
 and solves this equation in the limit $r -> \infty$ one finds
 
 \begin{equation}
 \rho = {\sigma^2 \over 2 \pi r^2 G}
 \end{equation}
 
 This configuration, called an isothermal sphere, involving gravitational collapse of collisionless
 particles strongly suggests that the dark matter does not interact strongly or electromagnetically.
 In addition, estimating the total dark matter around galaxies implies a lower bound of
 $\Omega_m > O(0.1)$, which exceeds the total baryonic matter density.  It is for this reason
 that cosmologists were initially driven to consider exotic non-baryonic dark matter.
 
 The next question, of course, is how much dark matter is there out there?
Perhaps the second greatest change in cosmological prejudice in the past decade
relates to the inferred total abundance of matter in the Universe. 
Because of the great intellectual attraction Inflation as a mechanism to
solve the so-called Horizon and Flatness problems in the Universe, it is
fair to say that most cosmologists, and essentially all particle
theorists had implicitly assumed that the Universe is flat, and thus that
the density of dark matter around galaxies and clusters of galaxies was
sufficient to yield $\Omega =1$.   Over the past decade it became more
and more difficult to defend this viewpoint against an increasing number
of observations that suggested this was not, in fact, the case in the
Universe in which we live.

The earliest holes in this picture arose from measurements of galaxy
clustering on large scales.  
  The growth of structure in the Universe, if gravity is responsible for such
growth, provides an excellent probe of the universal mass density, based largely
on issues associated with causality alone.   The basic idea is the following: 
If primordial density fluctuations have no preferred scale, then one can express
their Fourier transform as a simple power of the wavenumber $k$.  At the same
time, if this power is much greater than unity, density fluctuations will blow up
for large wavenumber, or small wavelength, and too many primordial black holes
will be created.  If the power is much less than unity, then fluctuations on
large scales (small wavenumbers) will be inconsistent with the observed isotropy
of the Cosmic Microwave Background radiation.   Thus, we expect the exponent,
$n$ to be near one, and inflationary models happen to predict precisely this
behavior.  

The primordial power spectrum, however, is not what we observe today, as
density fluctuations can be affected by causal microphysical processes once the
scale of these fluctuations is inside the horizon scale---the distance over
which light can have travelled between t=0 and the time in question.  One can
show that in an expanding universe, as long as the dominant form of energy
resides in radiation, gravity is ineffective at causing the growth of density
fluctuations.  In fact, such primordial fluctuations in baryons will be damped
out due to their coupling to the radiation gas.   Once the universe becomes
matter dominated, however, primordial fluctuations on scales smaller than the
horizon size can begin to grow.

\begin{figure}
\centerline{\psfig{figure=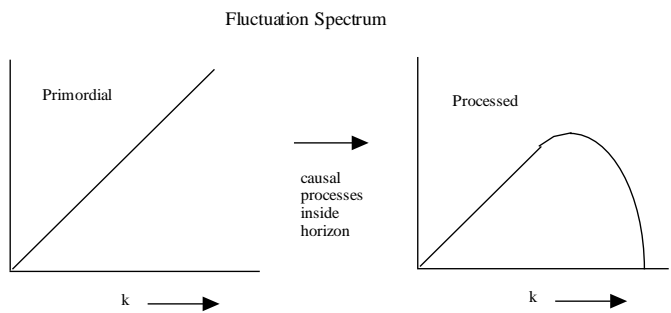,width=9.0cm}}
\label{puert1}
\caption{Processing of the initial power spectrum due to damping of high frequency
modes during radiation domination}
\end{figure}

These arguments suggest that an initial power law spectrum of fluctuations will
``turn over" as shown
in Figure 12  
for large wavenumbers which entered inside the causal horizon during
the early period of radiation domination in the Universe.  By exploring the
nature of the clustering of galaxies today over different scales, including
measurements of the two point correlation function of galaxies, the angular
correlation of galaxies across the sky on different scales, etc, one can hope to
probe the location of this turn-around, and from that probe the time, and
thus the scale which first entered the horizon when the universe became matter
dominated.  Clearly this time will depend upon the ratio of matter to radiation
in the Universe today (if this ratio is increased, then matter, whose density
decreases at a slower rate than radiation as the universe expands, will begin to
dominate the expansion at an earlier time, and vice versa. In turn, knowing this
ratio today gives us a handle on $\Omega_{\rm matter}$.

 Making the assumption that dark matter dominates on large
scales, and moreover that the dark matter is cold (i.e. became
non-relativistic when the temperature of the Universe was less than about
a keV), fits to the two point correlation function of galaxies on large
scales yielded (\cite{peac,liddle}):

\begin{equation}
\Omega_M h =.2-.3
\end{equation}

Unless $h$ was absurdly small, this would imply that $\Omega_M$ is
substantially less than 1.
  
New data from the Sloan and 2DF surveys refine this limit further, with reported values of (\cite{2df,sloan})

\begin{eqnarray}
\Omega_M=  0.23 \pm 0.09 (2DF) \\
\Omega_Mh  \approx 0.14^{+.11}_{.-06} \ (2\sigma)  \  (Sloan)
\end{eqnarray}

The second nail in the coffin arose when observations of the evolution
of large scale structure as a function of redshift began to be made.
Bahcall and collaborators (\cite{neta}) argued strongly that evidence for
any large clusters at high redshift would argue strongly against a flat
cold dark matter dominated universe, because in such a universe structure
continues to evolve with redshift up to the present time on large scales,
so that in order to be consistent with the observed structures at low
redshift, far less structure should be observed at high redshift.  Claims
were made that an upper limit $\Omega_B \le 0.5$ could be obtained by
such analyses.

A number of authors have questioned the systematics inherent in the early
claims, but it is certainly clear that there appears to be more structure
at high redshift than one would naively expect in a flat matter dominated
universe.  Future studies of X-ray clusters, and use of the
Sunyaev-Zeldovich effect to measure cluster properties should be able to
yield measurements which will allow a fine-scale distinction not just
between models with different overall dark matter densities, but also
models with the same overall value of $\Omega$ and different equations
of state for the dominant energy (\cite{mohr}).

One of the best overall constraint on the total
density of clustered matter in the universe comes from the combination of
X-Ray measurements of clusters with large hydrodynamic simulations.  The
idea is straightforward.  A measurement of both the temperature and
luminosity of the X-Rays coming from hot gas which dominates the total
baryon fraction in clusters can be inverted, under the assumption of
hydrostatic equilibrium of the gas in clusters, to obtain the underlying
gravitational potential of these systems.  In particular the ratio of
baryon to total mass of these systems can be derived.   Employing the
constraint on the total baryon density of the Universe coming from BBN,
and assuming that galaxy clusters provide a good mean estimate of the
total clustered mass in the Universe, one can then arrive at an allowed
range for the total mass density in the Universe
(\cite{white,krauss2,evrard}).  Many of the initial systematic
uncertainties in this analysis having to do with cluster modelling have
now been dealt with by better observations, and better simulations
( i.e. see \cite{mohr2}), so that now a combination of BBN and cluster
measurements yields:
\begin{equation}
\Omega_M = 0.35 \pm 0.1  \ \ (2\sigma)
\end{equation}

Combining these results, one derives the constraint:

\begin{equation}
\Omega_M \approx 0.3 \pm 0.05 \ \ (2\sigma)
\end{equation}

\subsubsection{Equation of State of Dominant Energy:}

The above estimate for $\Omega_M$ brings the
discussion of cosmological parameters full circle, with consistency
obtained for a flat 13 billion year old universe , but not one dominated
by matter.  As noted previously, a cosmological constant dominated 
universe with $\Omega_M = 0.3$ has an age which nicely fits in the
best-fit range.   However, based on the data discussed thus far, except for
the CMB data which is consistent with a flat universe dominated by
dark energy, there was
no direct evidence that the dark energy necessary to result in a flat
universe actually has the equation of state appropriate for a vacuum
energy.  Direct motivation for the possibility that the dominant energy
driving the expansion of the Universe violates the Strong Energy
Condition actually came somewhat earlier, in 1998, from two different sets of observations of
distant Type 1a Supernovae.  In measuring the distance-redshift relation
(\cite{perl,kirsh}) these groups both came to the same, surprising
conclusion:  the expansion of the Universe seems to be accelerating. 
This is only possible if the dominant energy is
"cosmological-constant-like", namely if "$\omega < -0.3$  (recall that
$\omega =-1$ for a cosmological constant).  

Note, as I 
have described, that the CMB data combined with supernova measurements now
favor a range $ -1.2 < w < -0.8$. 
In order to try and determine if the dominant dark energy does in fact
differ significantly from a static vacuum energy---as for example may
occur if some background field that is dynamically evolving (see next section) is
dominating the expansion energy at the moment---one can hope to search for
deviations from the distance-redshift relation for a cosmological
constant-dominated universe.  To date, none have been observed.  
Either other measurements,
such as galaxy cluster evolution observations, or space-based SN
observations would be required to further tighten the constraint.

\subsection{Summary}

I list the overall constraints on cosmological parameters discussed above
in the table below.  It is worth stressing how completely
remarkable the present situation is.  After 20 years, we now have the
first direct evidence that the Universe might be flat, but we also have
definitive evidence that there is not enough matter, including dark
matter, to make it so.  We seem to be forced to accept the possibility
that some weird form of dark energy is the dominant stuff in the
Universe.  It is fair to say that this situation is more mysterious, and
thus more exciting, than anyone had a right to expect it to be.

\begin{table}
\caption{Cosmological Parameters }
\begin{center}
\begin{tabular}{|l|l| c|}
\hline
Parameter & Allowed range  & Formal Conf. Level (where approp.) \\
\hline
 $H_0$ & $70 \pm 5$ & $ 2 \sigma$ \\
 $t_0$ & $13.7 \pm 0.2$ Gyr & $2 \sigma$ \\
 $\Omega_B h^2$ & $.02 \pm .004$ &$2 \sigma $ \\
 $\Omega_B$ & $0.045 \pm 0.015 $ & $2 \sigma $ \\ 
 $\Omega_M$ & $0.3 \pm 0.1 $ & $2 \sigma $ \\
 $\Omega_{TOT}$ & $1.02 \pm 0.04 $ & $2 \sigma $ \\
 $\Omega_{X}$ & $0.7 \pm 0.1 $ & $2 \sigma $ \\
 $\omega$ & $-1.2 - -0.8$ & $2 \sigma $ \\ \hline
\end{tabular}
\end{center}
\end{table}

\section{ Dark Matter and Dark Energy: A Particle Physics Perspective}

\subsection{Dark Matter}

The possible existence of non-baryonic dark matter should not come as a
surprise.  After all, while normal matter is all we are familiar with, by number baryons
are an almost insignificant fraction of the universe.  There is one baryon per billion
photons in the CMB, for example.   Moreover, the CMB remained hidden until 
1965, even though there is nothing more visible than electromagnetic radiation!  

Thus, it is easy to imagine how some particles could have been created in the
early universe with a remnant abundance far bigger than baryons, and which could still
remain unobserved.    In fact, virtually every single extension of the standard model of
particle physics predicts natural dark matter candidates, and as we shall see, even the
standard model itself includes particles which could have been fine dark matter 
candidates!!   The surprise, in this sense, 
would have been if no dark non-baryonic matter were discovered.

There are three different mechanisms by which elementary particle dark matter can be
created.  Some elementary particles are either:

\vspace{1pc}

1.  Born Dark!

\vspace{1pc}

2. Achieve Dark Matter-dom!

\vspace{1pc}

3. Have Dark Matter-dom thrust upon them!

\vspace{1pc}

\noindent{1. There are two examples of the first type: light neutrinos and monopoles.   Neutrinos
were the first non-baryonic dark matter candidate proposed because we know there is
a cosmic neutrino background of about 100 neutrinos/cc permeating space, just as there
is a cosmic microwave background of electromagnetic radiation.  Neutrinos were present
in thermal equilibrium in the early hot-dense phase of the universal expansion with an
abundance, per helicity state of:}

\begin{equation}
\zeta (3) {3 T^3 \over 4 \pi^2}
\end{equation}
where $\zeta$  is the Riemann zeta function.

One can solve Boltzmann's equation for neutrinos to determine when they go out of
thermal equilibrium.  A rough approximation is when their weak interaction rate is smaller
than the expansion rate.  Performing such a calculation for light neutrinos yields a decoupling
temperature of 2 MeV.   Since electrons and positrons annihilate with each other when the
universe drops below this temperature, and these particles dump their energy and entropy
into photons, the remnant neutrino temperature is reduced compared to photons by
a factor of $(4/11)^{1/3}$, which can be calculated by considering the number of helicity states in equilibrium in the radiation gas both before this entropy is dumped and afterwards.
Plugging in this number density today, one finds that if light neutrinos have a mass of approximately 10 eV, they could account for all the dark matter in the universe.   

As you can see, almost none of this depends upon detailed interactions of neutrinos---they just have to be initially part of the heat bath, then decouple.  Then, just by existing in the early universe, and then having a small mass, they would dominate the universe today! The same is true for magnetic monopoles, which can be produced as topological defects in the early universe.  If a phase transition occurs in which Grand Unified gauge group breaks to yield a U(1) subgroup, then roughly one defect will form per horizon volume at that time. Again, for the appropriate mass range, and assuming inflation does not occur after this transition, then monopoles could ultimately dominate the mass density of the universe today.  

\vspace{1pc}
\noindent{2.  Neutrinos also are the prototypes for the second kind of dark matter, the dark matter achievers.   In this case, the particles in question would be ultra-heavy neutrinos, greater than about
1 GeV.  These were the prototypical Weakly Interacting Dark Matter (WIMP) candidates.  }

In the case of heavy neutrinos, the weak interactions of these particles causes them to decouple from the heat bath only once the temperature has fallen far below their mass.  During the intervening period, neutrinos and antineutrinos annihilate, suppressing their number density relative to photons by
a factor $exp(-M/T)$.    In this case the details of their interactions are extremely important, because these details determine their decoupling temperature and therefore their remnant abundance. 

For WIMPs one finds a relatively general relation based on
incomplete annihilations in an initial thermal poulation:
\begin{equation}
\Omega_X h^2 \approx { 10^{-37} {\rm cm^2} \over <\sigma_{ann} v>}
\end{equation}

Considering a heavy Dirac neutrino with standard weak interactions,  one finds that a 2 GeV neutrino would just close the Universe today.   Interestingly, since the annihilation cross section increases with neutrino mass, heavier neutrinos would contribute a {\it smaller} relic density today.

As a result of experiments at LEP, and also searches for heavy cosmic neutrinos we know that no 
such heavy WIMP neutrinos exist.  Moreover, there was no reason to expect them to, as there is
nothing special about the GeV mass range for neutrinos in the standard model.  However, if we extend the  standard model to incorporate low energy supersymmetry breaking in order to to attempt to resolve 
the hierarchy problem, then one expects the masses of supersymmetric partners of ordinary matter will lie near the weak scale.   The lightest supersymmetric partner (LSP) is generally stable in most such models, and since it interacts with ordinary matter via the exchange of supersymmetric particles whose mass is on the order of the weak scale, the LSP is a natural version of a heavy neutrino WIMP!
Hundreds of calculations have been performed over the years, and there is significant phase space in
supersymmetric models for WIMP dark matter, having densities $\Omega \approx 0.1-0.3$.

What makes WIMPs so interesting is that because their remnant abundance is determined by
their annihilation cross section, one can use crossing symmetry to get a direct relation between this quantity, which determines their relic abundance, and their scattering cross section with matter.  
As a result of this, one can determine that these WIMPs may be detectible in direct detection 
experiments, as you shall read in this volume.   But also because of this, it is actually good news
that the density of WIMPs is {\it smaller} than we had previously imagined it to be.  
For both WIMPs, and for the other well motivated cold dark matter candidate that I have not yet
discussed, axions, one can write down a general
relation:

\begin{equation}
\sigma_{detection} \approx {1 \over \Omega_{DM}}
\end{equation}

The reasons for this are different for each candidate.  For WIMPs it is obvious. Because 
remnant abundance decreases as the annihilation cross section increases, and because of
the crossing symmetry relation between annihilation and scattering cross sections, 
as the WIMP
abundance decreases, its scattering cross section generally increases.

Astute experimentalists may argue that this is a scam, because as the
WIMP (axion) density decreases, the flux on Earth also decreases, so
even if there are larger cross sections, the event rate will not change!
However, this is wrong.  Until the density decreases to the point (below
about $\Omega_x < 0.1$ ) when WIMPs (axions) do not have sufficient
densities to account for all galactic halo dark matter, it is natural to
assume that their galactic density is given by the halo density.  Just
because their overall cosmic density is insufficient to close the universe,
this need not imply that their flux on earth is reduced.

\vspace{1pc}
\noindent{3. As advertised, axions are an example of the third class of `forced' dark matter
candidates.
Axions are pseudo-goldstone bosons that get a very small mass due to QCD effects in a way
which is associated with the solution of the Strong CP Problem.
Because they are goldstone bosons, axion fields can be represented as an angular field.
In spite of their very small mass, which would mean that any axions that are thermally produced 
in the early universe would provide a negligibly small contribution to the energy density 
today, a non-thermal production mechanism changes everything.}

  At early times their potential (considered as a function of an 
angular variable which can be taken to go from $-\pi$ to $\pi$) changes, as seen in figure 13.

\begin{figure}
\centerline{\psfig{figure=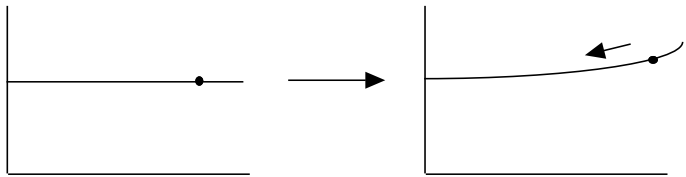,width=9.0cm}}
\label{peurt4}
\caption{Axion Potential as the Universe cools}
\end{figure}

In the former case, no energy is stored in the axion field.  However, 
once the axion gets a mass, energy is stored in the axion field, which
then dynamically rolls to the bottom of its potential.  However, the
time it takes to begin rolling is inversely proportional to the 
curvature of its potential, and is thus inversely proportional to the
axion mass.  Thus, the smaller the axion mass, the longer the energy
gets stored before it begins to redshift and the greater the remnant axion
density.  One finds that for an axion with mass approx. $10^{-5}$ eV,
the axion density can be naturally in the range of $\Omega =1$.

Axions too are detectable, in principle, because they have a coupling to
two photons.   Thus, even though their coupling is extremely small if the
symmetry breaking scale associated with their existence is large, they could
coherently couple to a large magnetic field within some volume, converting
into photons with a frequency equal to the axion mass.   Experiments designed
to detect such cosmic axion-photon conversion are currently underway.

By the way, because of this non-thermal
production mechanism, axions too share with WIMPs the fact that the smaller their
cosmic density today, as long as it is larger than the galactic halo density, 
the larger the cosmic axion interaction rate in detectors would be.   This is because axion couplings are inversely proportional to the
axion mass.

\subsection{Dark Energy}

The equation of state for dark energy that appears to be favored by the
existing data is $w=-1$, which is precisely that predicted for `vacuum' energy, 
which is in turn precisely that of a cosmological constant originally introduced
by Einstein into his equations when he thought the universe was static.  

Once relativity and quantum mechanics were combined, the existence of
vacuum energy was in some sense inevitable, as zero point energy is 
inevitably associated
with quantum mechanical ground state configurations.   Indeed, in the
current picture 
magnitude of the vacuum energy associated with a
cosmological constant which would be required by the present data is
remarkably small. 
After all, if it is quantum gravity at the Planck scale that cuts off the 
magnitude of quantum fluctuations, then the natural magnitude of the
vacuum energy density would be, by dimensional analysis, $M_{Pl}^4$.

 To get a sense of experimentally how remarkably small the energy 
 density associated with the dark energy today is by comparison, 
 consider the following experimental proposal:

\begin{center}
{\it ``To see what is in front of one's nose requires a constant struggle''}
\rightline{George Orwell}
\end{center}

What, you may ask, does this have to do with the topic at hand.  Plenty, I claim.
For it reminds us that we can put remarkably stringent limits on certain 
quantities by using macroscopic amounts of material.  In particular, it harkens
back to another famous quotation, this time from Maurice Goldhaber, who put
one of the first limits on proton decay by declaring that if the proton had
a lifetime less than about $10^{17}$ years, ``{\it You could feel it in your
bones!''}.  By this he meant that proton decays in our body would be so
frequent that we would die from the radiation exposure.

In this spirit we can perform a similar experiment.   Look at the end of your
nose.   Now, in a universe dominated by a cosmological constant, space begins
to expand exponentially.  One can calculate than for distances separated by
larger than an amount $R > M_{Pl}/3 \Lambda^{1/2}$, points will have a relative
velocity exceeding that of light, and thus will remain out of causal contact.
Thus, the fact that you can see the end of your nose implies a bound
$\Lambda < 10^{-68} M_{Pl}^4$!

Of course, the fact that we can see distant galaxies gives us an even stronger
bound.  And, the fact that the cosmological constant affects dynamics on larger
scales no more than it is claimed to by the present observations gives a bound
$\Lambda < 10^{-123} M_{Pl}^4$.   What makes this small number so hard to 
understand, in a cosmological context is not merely the ``naturalness'' problem
of which particle physicists are aware, but rather, if this has been constant
over cosmological time, this is the first time in the history of the universe
when the energy density in a cosmological constant is comparable to the
energy density of matter and radiation!  It is for this reason that some
cosmologists are driven to the idea that what is being observed is not really
a cosmological constant, but something perhaps more exotic (e.g. \cite{stein}).

There is a problem, from my point of view, however, with all of these proposals,
which is why I have  publicly bet my house on the fact that observers are
bound to measure $w=1$ when all is said and done.   The reasoning is quite
simple.  One can imagine a background scalar field, $\phi$, rolling down to the minimum
of its potential.  If it has not yet settled at its minimum, then the equation of state
for the field is given by:

\begin{equation}
\nonumber
w={{\dot\phi^2/2 -V} \over {\dot\phi^2/2 +V } }
\end{equation}

Since the kinetic energy of the scalar field as it rolls in the potential gives a positive contribution to the pressure, any rolling implies $w > -1$.     

At some level, most models of a non-cosmological constant type of dark energy rely on such
a mechanism to produce equations of state close to, but not equal to that of a cosmological
constant.   The problem is that in order to have a curvature so small so that the field has not
yet reached the minimum of its potential, one generally is required to have extremely small
masses associated with the field.   But beyond this fine tuning problem,
there are three problems I see with this:

(i) if the field potential can be fine tuned to take $10^{10}$ years to begin to roll down its potential,
then why not imagine a field that takes $10^{100}$ years to do so?    Indeed, it seems to me
that for such a field to just begin rolling now is HIGHLY coincidental and contrived.  If this were not the  case, the field, stuck at a non-zero value of its potential now, would be observationally indistinguishable from a cosmological constant. 

(ii)  What about the cosmological constant problem?  If this wierd field is to dominate the energy of empty space today, then somehow vacuum fluctuations must give a yet much smaller, or zero contribution.  But this requires solving the cosmological constant problem anyway.

(iii)  Finally, we should remember that almost all quantum field theories PREDICT a non-zero cosmological constant.  The only problem is making it small enough to agree with observations.  Thus, a cosmological constant is, in a sense, the most natural candidate for dark energy.  All we have to do is figure out why...

\section{Geometry, Destiny, and the Future}

Once we accept that we live in a unvierse dominated
by dark energy, everything about the way we think about
cosmology changes.  In the first place, Geometry and Destiny
are no longer linked.  Previously, the holy grail of cosmology involved
determining the density parameter $\Omega$, because this was tantamount
to determining the ultimate future of our universe.  Now, once we accept
the possibility of a non-zero cosmological constant, we must also accept
the fact that any universe, open, closed, or flat, can either expand
forever, or reverse the present expansion and end in a big crunch
(\cite{kraussturn}).  

{}The mathematical basis of this is described simply.  Einstein's equations
imply, for an isotropic and homogeneous Universe, the following evolution
equations for the cosmic scale factor, $R(t)$:
\begin{eqnarray}
 H^2\equiv { \left ( \dot R \over R \right )}^2 & = &
 {8 \pi G\over 3}  \rho_{\rm TOT}  - {k \over R^2}
\label{eq:rdot}  \\
{\ddot R  \over R} & = & -{4 \pi G\over 3}\sum_i\rho_i(1 + 3w_i)
\label{eq:rdotdot}
\end{eqnarray}
where $k$ is the signature of the 3-curvature,
the pressure in component $i$ is related to the energy
density by $p_i = w_i \rho_i$ and the total energy density
$\rho_{\rm TOT} = \sum_i\rho_i$.
The evolution of the energy density in component $i$ is determined by
\begin{equation}
{d\rho_i \over \rho_i} = -3(1+w_i){dR\over R}\ \
\Rightarrow \ \ \rho_i\ \propto\ R^{-3(1+w_i)}
\label{eq:firstlaw}
\end{equation}

{}All forms of normal matter satisfy the strong-energy condition,
$(\rho_i  + 3p_i)=\rho_i(1+3w_i)>0$, and so if
the Universe is comprised of normal matter,
the expansion of the Universe always decelerates, cf. Eq.~(\ref{eq:rdotdot}).
Also, since $\rho$ is positive for normal matter,
the first equation implies that $ \dot R /R$
remains positive and non-zero if $k \le 0$, and thus the
Universe expands forever.  Equation (\ref{eq:firstlaw}) and
the strong-energy condition imply that $\rho_i$ decreases more
rapidly than $R^{-2}$.  Thus, for $k>0$ there is necessarily
a turning point with $H=0$ and $\ddot R < 0$, and the Universe
must ultimately recollapse.  Geometry determines destiny.

{}However, a cosmological constant
violates the strong-energy condition, completely obviating the logic of the above
argument.   Recalling that  $p_\Lambda =-\rho_\Lambda$ for a
cosmological term, and that $p_M=0$
for matter, the above equations become,
\begin{eqnarray}
H^2 & = & {8 \pi G\over 3}  (\rho_M + \rho_{\Lambda})  - {k \over R^2}
\label{eq:rdot_lambda} \\
{\ddot R \over R} & = & -{4 \pi G\over 3}(\rho_M -2 \rho_{\Lambda})
\label{eq:rdotdot_lambda}
\end{eqnarray}

{}Since $\rho_{\Lambda} = $ constant, while $\rho_M \propto R^{-3}$,
even if $k >0$, as long as $H > 0 $ when $\rho_{\Lambda}$
comes to dominate the expansion, it will remain positive forever,
and as is well known, the expansion will ultimately accelerate,
$R(t) \rightarrow e^{Ht}$ with $H=\sqrt{8\pi G\rho_\Lambda /3}$.

{}One conventionally defines the scaled energy density
$\Omega \equiv \rho_{\rm TOT} /\rho_{\rm crit} =
8 \pi G \rho /3H^2$, so that $\Omega -1 = k/H^2R^2$.
Thus the sign of $k$ is determined by whether $\Omega$ is
greater than or less than 1.
In this way, a measurement of $\Omega$ at any epoch -- including
the present -- determines the geometry of the Universe.  However,
we can no longer claim that the magnitude of $\Omega$ uniquely
determines the fate of the Universe.

It is interesting to determine how
small a cosmological constant could be at the present time and still stop the
eventual collapse of a closed Universe.  For a closed, matter-only Universe,
the scale factor at turnaround is
\begin{equation}
{R/R_0} = {\Omega_0 \over \Omega_0 - 1}
\end{equation}
While all the evidence today suggests
that $\Omega_0 \le 1$, existing uncertainties could allow $\Omega_0$ to
be as large say as 1.1.  For $\Omega_0=1.1$ the scale
factor at turnaround is $11R_0$.   Since the density of matter decreases as
$R^{-3}$, this means that an energy density in a cosmological term as
small as $1/1000th$ the present matter density will come to dominate the
expansion before turnaround and prevent forever recollapse.
A cosmological constant this small,
corresponding to $\Omega_\Lambda \sim 0.001$, is completely
undetectable by present, or foreseeable observational probes.

{}Alternatively, it may seem that if we can unambiguously determine that
$k <0$ then we are assured the Universe will expand forever.   However, this is
the case only as long as the cosmological constant is positive.  Since we have
no theory for a cosmological constant,
there is no reason to suppose that this
must be the case.   When the cosmological constant is negative, the energy
density associated with the vacuum is constant and {\it negative}.   In this
case, from Eqs.~(\ref{eq:rdot_lambda},\ref{eq:rdotdot_lambda}),
one can see that not only is the ultimate expansion
guaranteed to decelerate, but recollapse is also inevitable,
{\it no matter how small the absolute value of $\Omega_\Lambda$ is}.

{}Finally, what if we indeed ultimately verify $w=1$
at the present
time, as current observations suggest?  Even in this case we are not guaranteed an eternal
expansion.   As I have described  scalar field
which is not at the minimum of its potential will, as long as the age of the
Universe is small compared to the characteristic time it takes for the field to
evolve in its potential, mimic a cosmological term in Einstein's equations. 
Until the field evolves to its ultimate minimum, we cannot derive the
asymptotic solution of these equations in order to determine our destiny.

As Michael
Turner and I have demonstrated, there is no set of cosmological
measurements, no matter how precise, that will allow us to determine the
ultimate future of the Universe.  In order to do so, we would require a
theory of everything.   

On the other hand, if our universe is in fact dominated by a cosmological
constant, the future for life is rather bleak (\cite{kraussstark}). 
Distant galaxies will soon blink out of sight, and the Universe will
become cold and dark, and uninhabitable.  

This bleak picture may seem depressing, but the flip side of all the
above is that we live in exciting times now, when mysteries abound.  
We should enjoy our brief moment in the Sun.

\end{document}